\documentclass[a4paper,11pt]{article}
\pdfoutput=1 % if your are submitting a pdflatex (i.e. if you have images in pdf, png or jpg format)

\usepackage{jheppub} % for details on the use of the package, please see the JHEP-author-manual
\usepackage[T1]{fontenc} % if needed

\title{\boldmath ReSyst: a novel technique to Reduce the Systematic uncertainty for precision measurements}

\author[1]{P.~Van Mulders\note{Postdoctoral fellow and part-time (10\%) professor looking for a permanent position. Particularly interested in vacancies with the potential to solve the two-body problem.}}
\affiliation{Vrije Universiteit Brussel, Belgium}

\emailAdd{Petra.Van.Mulders@vub.be}

\abstract{We are in an era of precision measurements at the Large Hadron Collider. The precision that can be achieved on some of those is limited however due to large systematic uncertainties. This paper introduces a new technique to reduce the total systematic uncertainty by quantifying the systematic impact of single events and correlating it with event observables to identify classes of events that are more sensitive to systematic effects. A proof of concept is presented by means of a simplified top quark mass estimator applied on simulated events. Even without a thorough optimization, it is shown that the total systematic uncertainty can be reduced by at least 30\%.}

\newcommand{\ttbar}{$\mathrm{t \bar t}$}
\newcommand{\pt}{$p_\mathrm{T}$}
\newcommand{\mt}{$m_\mathrm{t}$}
\newcommand{\mjjj}{$m_\mathrm{jjj}$}

\begin{document} 
\maketitle
\flushbottom

\section{Introduction}
\label{sec:intro}
%Set the scene. 
With the large amount of data collected each year, the Large Hadron Collider (LHC) at CERN has entered an era of precision measurements. For the measurement of some of the standard model parameters, the systematic uncertainties largely limit the achievable precision. This is for instance the case for the most precise measurement of the top quark mass in a single decay channel, where the systematic uncertainty is already eight times larger than the statical uncertainty, i.e. $m_t = 172.25 \pm 0.08 \pm 0.62$~GeV~\cite{Sirunyan:2018gqx}. Over the next five years, the measurements at the LHC will be performed with up to ten times more data, reducing the statistical uncertainty on the most precise measurements to a negligible level. Therefore, the focus of the community is to reduce the systematic uncertainty for the precision measurement of the top quark mass. The ATLAS and CMS collaborations expect to reduce the dominant systematic uncertainties due to the jet energy scale, the b-quark fragmentation and the modelling of {\ttbar} events in the simulation~\cite{FTR-16-006}, and therefore to achieve a precision of around 300~MeV on the top quark mass at the end of the LHC Run 3. 

%Why this paper? 
When measurements in high-energy physics are dominated by systematic uncertainties, diverse techniques can be developed to reduce those, e.g. by tuning the event or reconstruction requirements~\cite{Abdallah:2008ad,BEDDALL2005469}. Also in the context of classification problems and machine learning some techniques have been developed to reduce the impact of sources of systematic uncertainties~\cite{LLyons,Baldi2016,Louppe2016}. In this article, a novel technique is presented to trade statistical precision for a reduced systematic uncertainty. The here introduced ReSyst technique can as well be considered complementary to existing efforts to reduce the total uncertainty. The ReSyst method revolves around the definition of a non-observable quantifier of the systematic impact for each event. Using simulation this non-observable quantifier can be correlated with event observables in order to identify classes of events inducing a relatively large impact in the total systematic uncertainty. The systematic uncertainty can then be reduced by rejecting events in specific parts of the phase space.

%Simulation
The concept is illustrated with a simplified top quark mass estimator using simulated {\ttbar} events at the LHC. The muon$+$jets decay channel is used, i.e. with one of the W bosons from the top quarks decaying to quarks (hadronic leg) and the other one to a muon and corresponding neutrino (leptonic leg). In total 10~M of these events are generated in proton-proton collisions at a centre of mass energy of 13~TeV, corresponding to about 80~fb$^{-1}$ of integrated luminosity, using the POWHEG v2 event generator~\cite{NLOQCD,POWHEG,POWHEGBOX,POWHEGhvq} and with the top quark mass set to 172.5~GeV. The generated events are interfaced with PYTHIA 8.2~\cite{PYTHIA} for the parton showering, hadronization and particle decay using the underlying event tune CUETP8M2T4~\cite{TOP-16-021} and further processed with DELPHES v3.4.2pre03~\cite{delphespaper,delphes1,delphes2} to simulate the CMS detector response. The anti-kt jet clustering algorithm with a distance parameter of 0.5~\cite{AntiKt} in the FastJet package~\cite{Cacciari:2011ma,hep-ph/0512210} is used to reconstruct jets when running DELPHES. The default DELPHES CMS parameter card is used with the exception of the b-tagging efficiency and misidentification probabilities for which the parametrizations for the medium working point of the DeepCSV algorithm developed by the CMS Collaboration are used~\cite{CMSbtagging}.

\section{Top quark mass estimator}
\label{sec:topmassest}
%Event selection and efficiency
Events are selected when they have at least one muon with transverse momentum (\pt) above 25~GeV within the CMS tracker acceptance ($|\eta|<2.4$) and at least four jets with \pt~$>30$~GeV within the tracker acceptance. At least two of the jets passing the {\pt} and $\eta$ requirements should be identified as originating from b quarks. The selection efficiency for the signal {\ttbar} events in the muon$+$jets decay channel is around 15\%. Other {\ttbar} decays or background processes are not considered in the analysis, which is motivated by the fact that the modelling of the background does not induce a dominant systematic uncertainty on the top quark mass~\cite{Sirunyan:2018gqx}. Moreover, the selection criteria which are applied reduce the fraction of background events to less than 10\% of the number of selected events~\cite{Sirunyan:2018gqx}. 

\begin{figure}[tbp]
\centering 
\includegraphics[width=.49\textwidth]{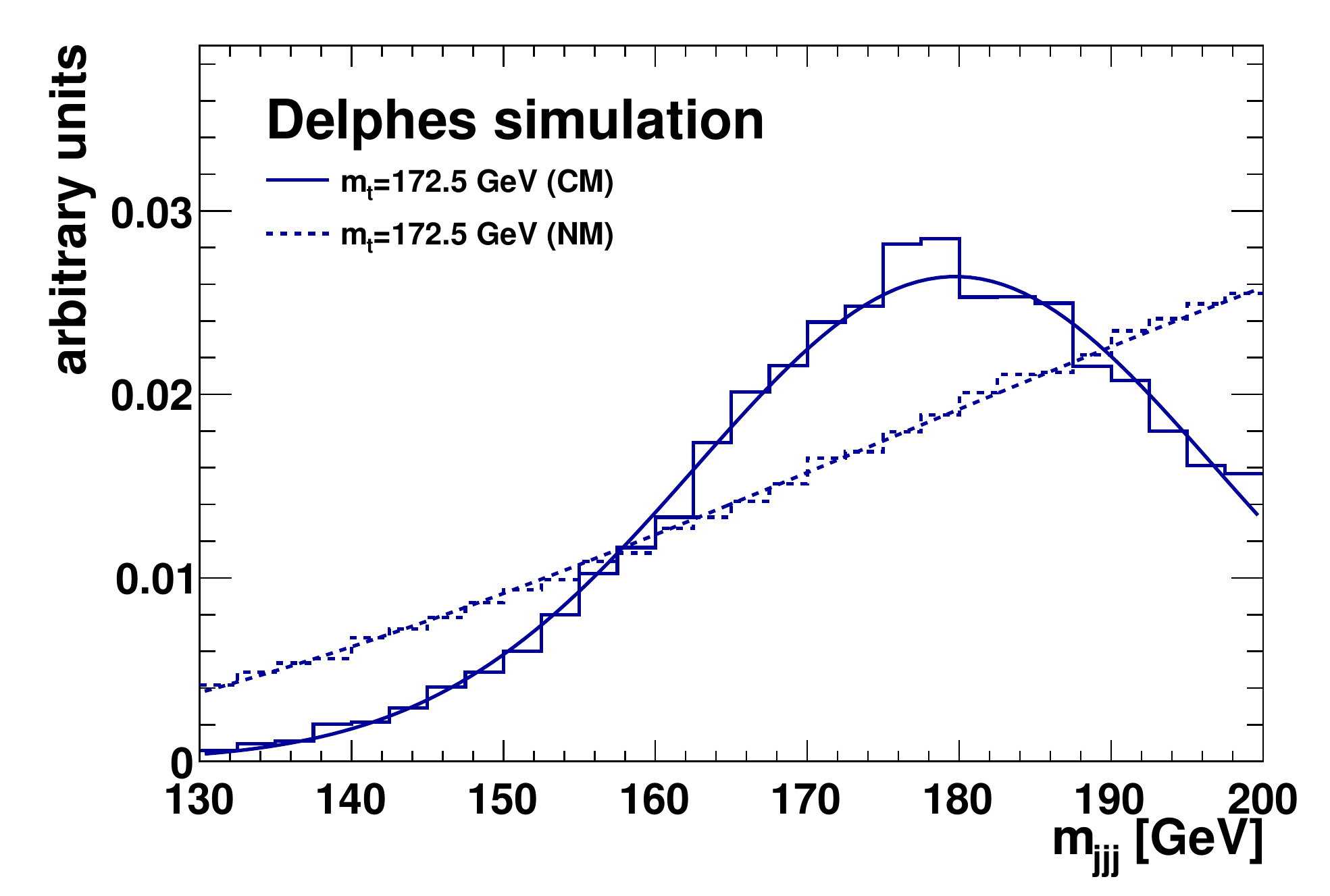}
\hfill
\includegraphics[width=.49\textwidth]{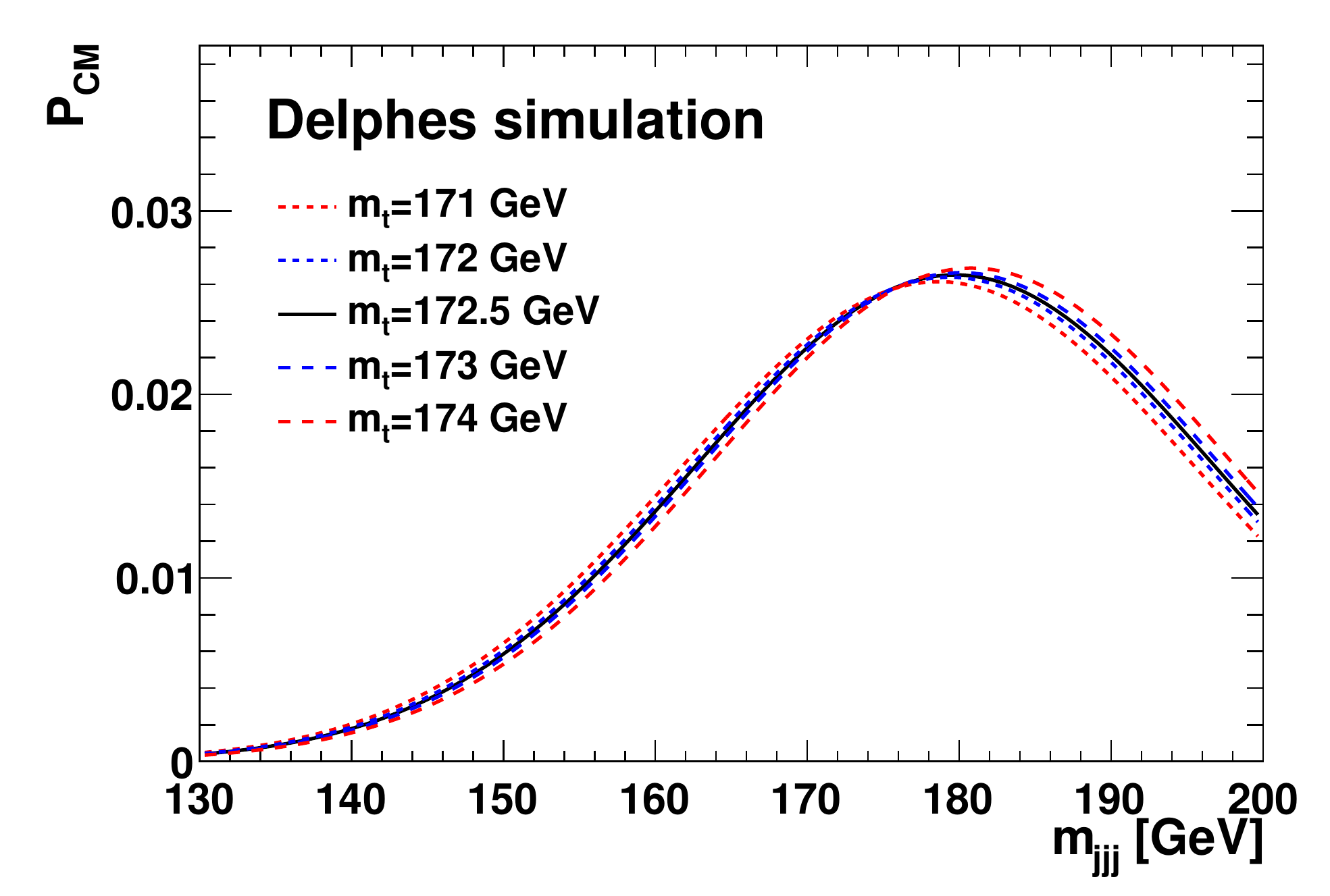}
\includegraphics[width=.49\textwidth]{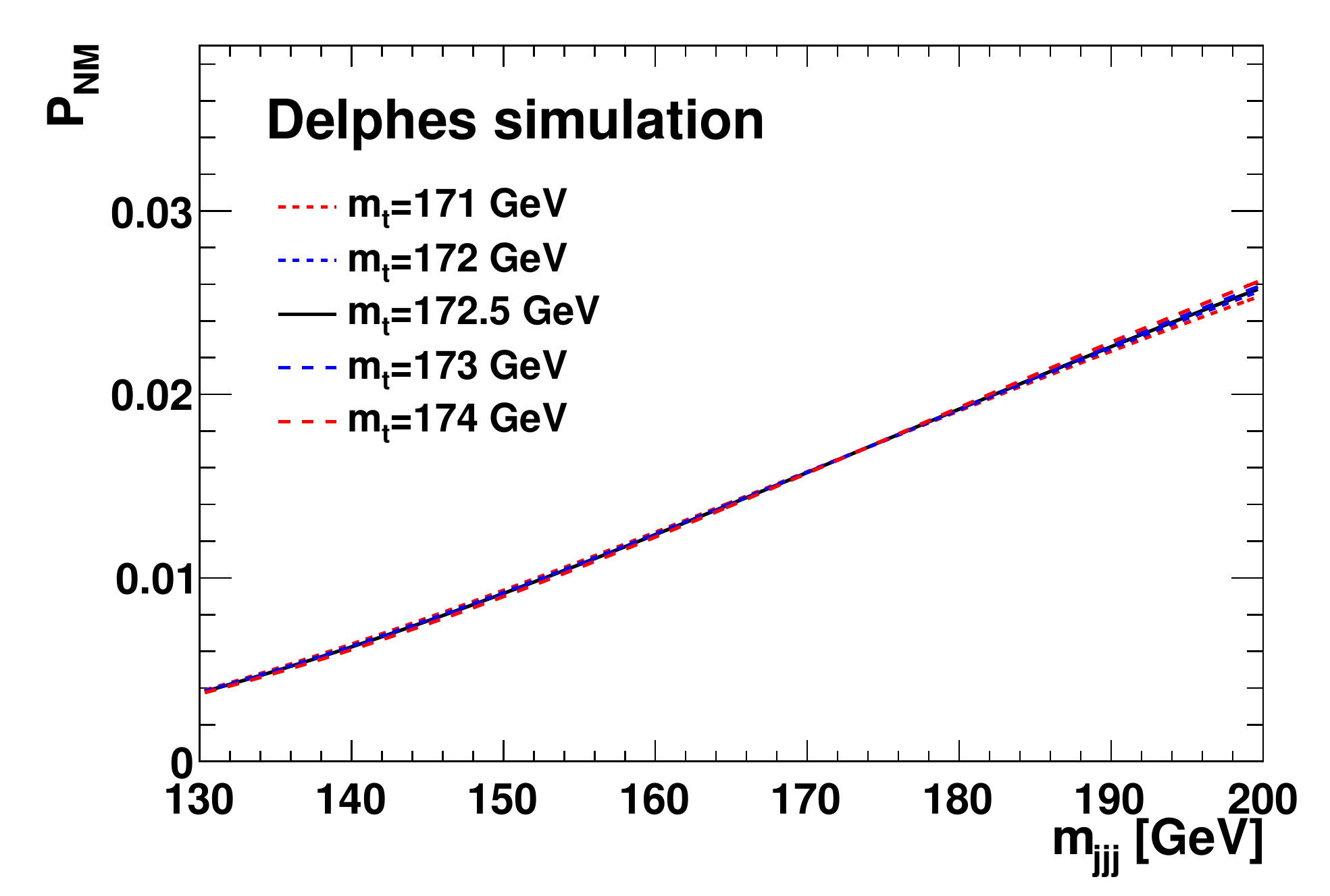}
\hfill
\includegraphics[width=.49\textwidth]{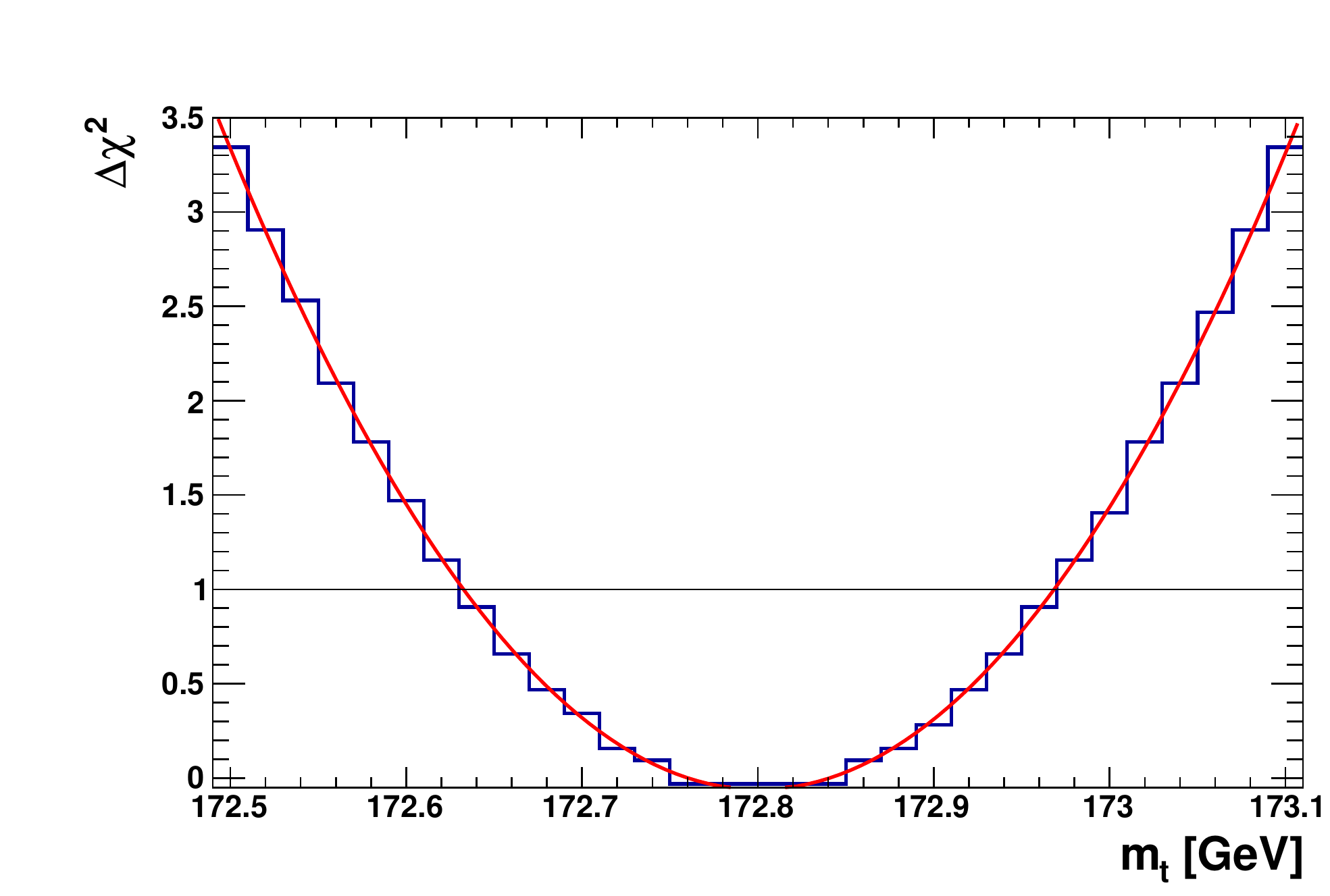}
\caption{\label{fig:templates} The {\mjjj} distribution for correctly and wrongly matched jet-quark combinations including the gaussian and third order polynomial functions fitted to the respective distributions (upper left). Probability density functions for correctly (upper right) and wrongly (down left) matched jet-quark combinations for different {\mt} values. The minimum of the fitted $\Delta\chi^2$ distribution corresponds to the estimated top quark mass (down right). A deviation of about 0.3~GeV is found compared to the generated mass. Usually a calibration is applied to correct this statistical bias.}
\end{figure}

%Top quark mass variable
The three highest-{\pt} jets are used to reconstruct the top quark corresponding to the hadronic leg of the {\ttbar} decay and its mass, {\mjjj}. To reduce the contribution of wrongly matched jet-quark combinations, i.e. those combinations not corresponding to the top quark decay, only events which have an {\mjjj} value between 130 and 200~GeV are considered. About 8\% of the initially selected events survive this additional requirement, which means that the total event selection efficiency drops to 1.2\%. To obtain samples with a different generated top quark mass, the events generated with {\mt} = 172.5~GeV are reweighted. The {\mjjj} distributions for correctly and wrongly matched jet-quark combinations, figure~\ref{fig:templates} are fitted with gaussian and third order polynomial functions, respectively, for different {\mt} hypotheses. The dependence of the parameters in the fitted functions on {\mt} is then fitted with first order polynomial functions. The obtained {\mt} dependence of the parameters is then used to construct the probability density functions $P_{CM}$ and $P_{NM}$ respectively for the correctly and wrongly matched jet-quark combinations. This procedure reduces the impact of sample specific statistical fluctuations in the {\mjjj} distribution and the chosen binning. The probability density functions for the selected events are used to construct a likelihood:
\begin{equation}
\label{eq:likelihood}
\mathcal{L}(m_\mathrm{t}) = \prod\limits_{i=1}^{n} f_{CM}(m_{t})P_{CM}(m_{\mathrm{jjj},i}|m_\mathrm{t}) + (1-f_{CM}(m_{t}))P_{NM}(m_{\mathrm{jjj},i}|m_\mathrm{t}),
\end{equation}
where the product runs over the number of events and the function $f_{CM}$ is the fraction of correctly matched jet-quark combinations corresponding to about 20.5\% in the considered {\mjjj} range. The functions $P_{CM}(m_{\mathrm{jjj},i}|m_\mathrm{t})$ and $P_{NM}(m_{\mathrm{jjj},i}|m_\mathrm{t})$ are the probability density functions for correctly and wrongly matched jet-quark combinations, respectively. The upper right (lower left) panel in figure~\ref{fig:templates} shows $P_{CM}(m_{\mathrm{jjj},i}|m_\mathrm{t})$ ($P_{NM}(m_{\mathrm{jjj},i}|m_\mathrm{t})$) for different {\mt} values. The dependence of $P_{NM}$ on the generated mass, {\mt}, is negligible except at high {\mjjj} values, which is an additional motivation to not consider events with {\mjjj} values above 200 GeV. The estimated top quark mass is obtained by evaluating the likelihood for different values of the generated top quark mass {\mt} and performing a maximum likelihood fit or minimizing $\Delta\chi^2=-2~\mathrm{ln}(\mathcal{L}(m_\mathrm{t}))$. The $\Delta\chi^2$ distribution is shown in the lower right panel of figure~\ref{fig:templates}. The minimum of the fitted function corresponds to the estimated top quark mass and the intersections of the horizontal line with the fitted function correspond to the size of the statistical uncertainty on the estimation.

%Systematic uncertainties
To illustrate the ReSyst technique for reducing the systematic uncertainties, the following systematic uncertainties are considered:
\begin{itemize}
\item {\bf b tagging efficiency and mistagging probability}: The uncertainty in the b tagging efficiency is typically 2\%, while the uncertainty in the mistagging probability is 5\% for c jets and up to 15\% for light-quark jets~\cite{CMSbtagging}. These uncertainties are taken into account by reweighting the events taking into account an independent variation upwards or downwards for the b tagging efficiency and misidentification probability using the true jet flavour. The difference between the estimated top quark mass for these variations and the nominal estimated top quark mass is taken as the systematic uncertainty. The square root of the quadratic sum of the systematic uncertainties for the b tagging efficiency and mistagging probability is used as the total uncertainty due to b tagging.
\item {\bf Jet energy scale}: The jet energy scale is varied upwards and downwards using a $p_{\mathrm{T}}$ and $\eta$ dependent variation corresponding to the jet energy scale uncertainty in~\cite{CMSJES}. The jet four-momentum is accordingly rescaled prior to the event selection.
\item {\bf Factorization and renormalization scales}: The factorization and renormalization scales ($Q^2$) at the matrix-element level are varied independently upwards and downwards with a factor of two. This gives rise to eight possible variations. The two variations where the factorization and renormalization scales are varied in opposite directions are unphysical and are therefore not considered. For the six remaining variations, the envelope is calculated to obtain the size of the systematic effect.
\item {\bf Matching between the matrix element and parton shower}: The matching between the matrix-element level and the parton shower is controlled by the so-called $h_{\mathrm{damp}}$ parameter. Radiated quarks and gluons are damped by a factor $h_{\mathrm{damp}}^2/(p_{\mathrm{T}}^2 + h_{\mathrm{damp}}^2)$. The parameter value was tuned to $1.581^{+0.658}_{-0.585}\times$~{\mt}~\cite{TOP-16-021}. The systematic uncertainties correspond to the difference between the estimated top quark mass when varying $h_{\mathrm{damp}}$ by the upward and downward uncertainty with respect to the nominal estimated value.
\item {\bf Top quark {\pt}}: The top quark {\pt} spectrum in data is observed to be softer than in simulated {\ttbar} events~\cite{TOP-16-011,TOP-16-008,TOP-12-028,TOP-14-018}. Therefore, the systematic effect due to the softer top quark {\pt} spectrum is taken into account by reweighting the {\pt} spectra of the two top quarks in the simulation. The difference with the top quark mass measurement before reweighting is taken as the size of the systematic effect. 
\item {\bf b quark fragmentation}: Another source of uncertainty is the modelling of the momentum transfer from the b quark to the B hadron during the parton shower. To assess the size of this systematic effect, the ratio of the {\pt} of the generated B hadron and the {\pt} of the b jet, $p_\mathrm{T}(B)/p_\mathrm{T}(\mathrm{b~jet})$, is varied by 2.5\% upwards and downwards. This number is motivated by the uncertainty on the $r_b$ parameter in the Lund-Bowler function which has been measured using e$^+$e$^-$ collisions from LEP and SLC~\cite{LEPrb,SLCrb} and which results in a variation of around 2.5\% in the $p_\mathrm{T}(B)/p_\mathrm{T}(\mathrm{b~jet})$ distribution. The top quark mass is remeasured for these variations and the difference with the nominal estimated value is quoted as the uncertainty.
\end{itemize}
When assessing the systematic uncertainty induced by each source the likelihood function in equation~\ref{eq:likelihood} remains unchanged. The sources of systematic uncertainties listed above are considered because they typically dominate the total uncertainty in top quark mass measurements. In addition, those uncertainties can be determined by reweighting events, avoiding the need to simulate additional event samples. The sources of systematic effects and their corresponding uncertainties are listed in table~\ref{tab:syst}. The systematic uncertainties due to the jet energy scale and the b quark fragmentation clearly dominate.
\begin{table}[tbp]
\centering
\begin{tabular}{|l|c|c|}
\hline
Systematic source & $+1\sigma$ effect [GeV] & $-1\sigma$ effect [GeV]\\
\hline
b tagging efficiency and mistagging probability & 0.01 & -0.01 \\
Jet energy scale & 0.88 & -0.87 \\ 
Factorization and renormalization scales & 0.01 & -0.02 \\
Matrix element and parton shower matching ($h_{\mathrm{damp}}$) & 0.04 & -0.01 \\
Top quark {\pt} & n.a. & -0.01 \\
b quark fragmentation & 0.39 & -0.41 \\
\hline
Total systematic uncertainty & 0.96 & -0.97 \\
\hline
\end{tabular}
\caption{\label{tab:syst}The sources of systematic uncertainties considered in the analysis and their impact on the top quark mass measurement. The jet energy scale and the b quark fragmentation dominate the total uncertainty.}
\end{table}
The size of the systematic uncertainties for the various sources are around the values expected for the 1D measurement discussed in Ref.~\cite{Sirunyan:2018gqx} with the exception of the b quark fragmentation modelling, which is larger in this document. However in the 1D measurement, an additional uncertainty is taken into account for the energy scale of b jets. The combination of the b jet energy scale uncertainty and the b quark fragmentation uncertainty in~\cite{Sirunyan:2018gqx} is in the same ballpark as presented here under the label b quark fragmentation. The b jet energy scale uncertainty and the b quark fragmentation uncertainty both account for the possible uncertainty in the energy of the b jet. Therefore, it is concluded that the impact of the systematic uncertainties on the estimated top quark mass in this case study is reasonable compared to the case presented in~\cite{Sirunyan:2018gqx}. Possible differences are related to the detector simulation and the choice of the top quark estimator.

%Result
The estimated mass is found to be $172.80 \pm 0.16~(\mathrm{stat.}) +0.96 -0.97~(\mathrm{syst.})$~GeV. The estimated mass is found to be 0.3~GeV different from the expected or generated mass of 172.5~GeV. Usually a calibration (or bias correction) is performed when estimating the top quark mass, e.g. as is done in Ref.~\cite{Sirunyan:2018gqx}, but in the context of this paper such a calibration is less relevant.

\section{Systematic effect quantifier}
\label{sec:quantifier}
%The new concept
The estimator presented in the previous section has a large systematic uncertainty, which is dominated by the uncertainty in the b quark fragmentation and the jet energy scale. If the impact of each event on the total systematic uncertainty could be quantified, we could reduce the systematic uncertainty by rejecting events inducing a large systematic effect. The ReSyst method revolves around quantifying the impact of an event $i$ on the total systematic uncertainty, denoted by $R_i$, as:
\begin{equation}
\label{eq:quantifier}
R_i = \frac{\sqrt{\sum\limits_{j}(m_{t(i)}^{+1\sigma_j}-m_{t(i)}^{-1\sigma_j})^2}}{\sqrt{\sum\limits_{j}(m_{t}^{+1\sigma_j}-m_{t}^{-1\sigma_j})^2}}.
\end{equation}
In this expression, the sum runs over all systematic sources $j$, $m_{t}^{+1\sigma_j}$ ($m_{t}^{-1\sigma_j}$) is the estimated top quark mass for a $+1\sigma$ ($-1\sigma$) variation of systematic source $j$, and $m_{t(i)}^{\pm 1\sigma_j}$ is the same but without considering event $i$. The difference $m_{t}^{+1\sigma_j}-m_{t}^{-1\sigma_j}$ quantifies the total effect of the upward and downward variation of the systematic source, while $m_{t(i)}^{+1\sigma_j}-m_{t(i)}^{-1\sigma_j}$ quantifies the effect without event $i$. In case the effect of a systematic source on the measurement is evaluated by a single variation, as is the case for the uncertainty on the generated top quark \pt, the difference is taken between this single variation and the nominal estimated top quark mass value, i.e. $m_{t}^{+1\sigma_j}-m_{t}$ and $m_{t(i)}^{+1\sigma_j}-m_{t(i)}$ without event $i$. The quantifier is inspired on the jackknife delete-1 resampling technique that can be used to estimate the variance and statistical bias of a measurement~\cite{jackknife}. In the case of the quantifier $R_i$ the systematic impact of a specific event is estimated by removing that event from the sample and repeating the estimation of the systematic uncertainties.

\begin{figure}[tbp]
\centering 
\includegraphics[width=.45\textwidth]{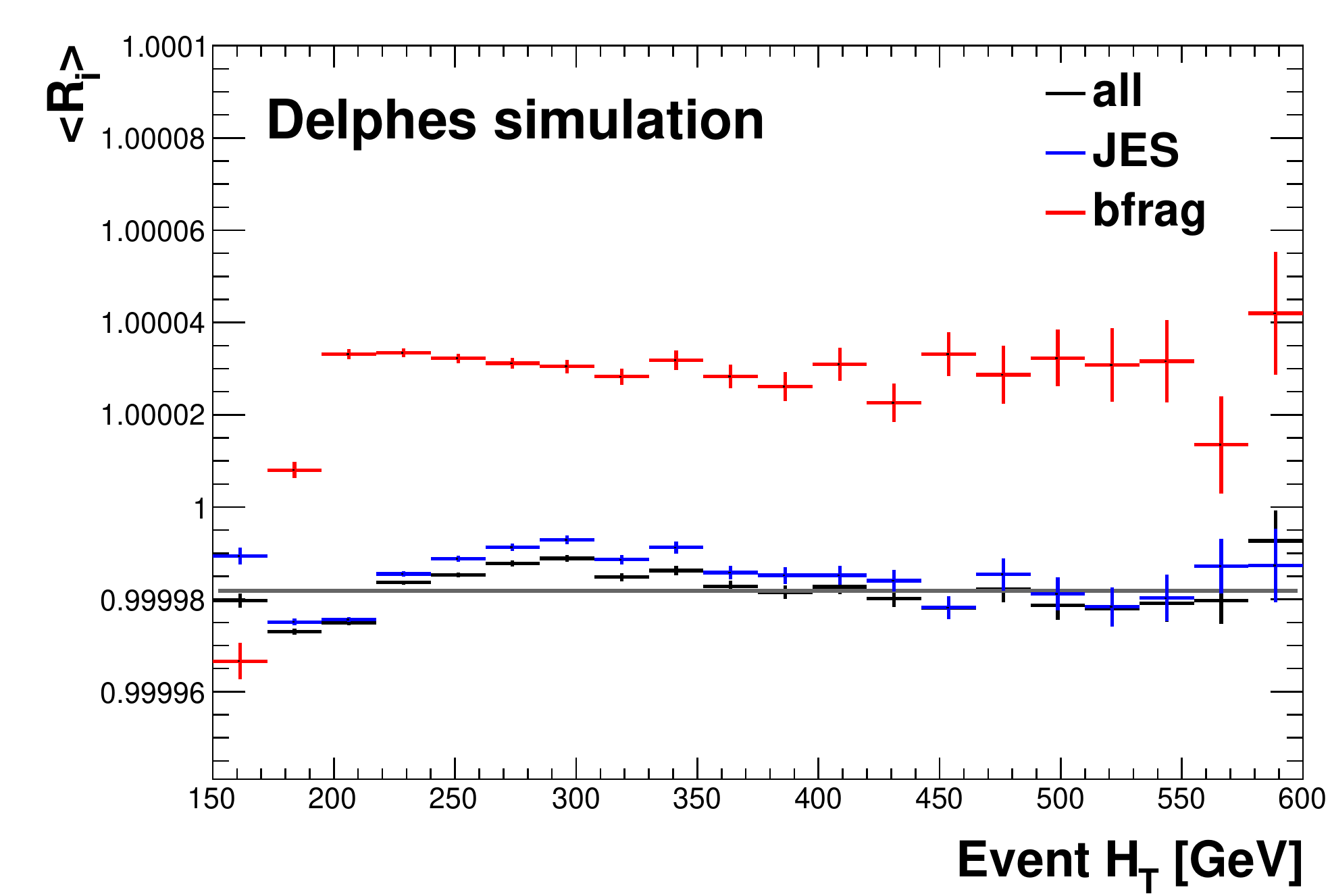}
\hfill
\includegraphics[width=.45\textwidth]{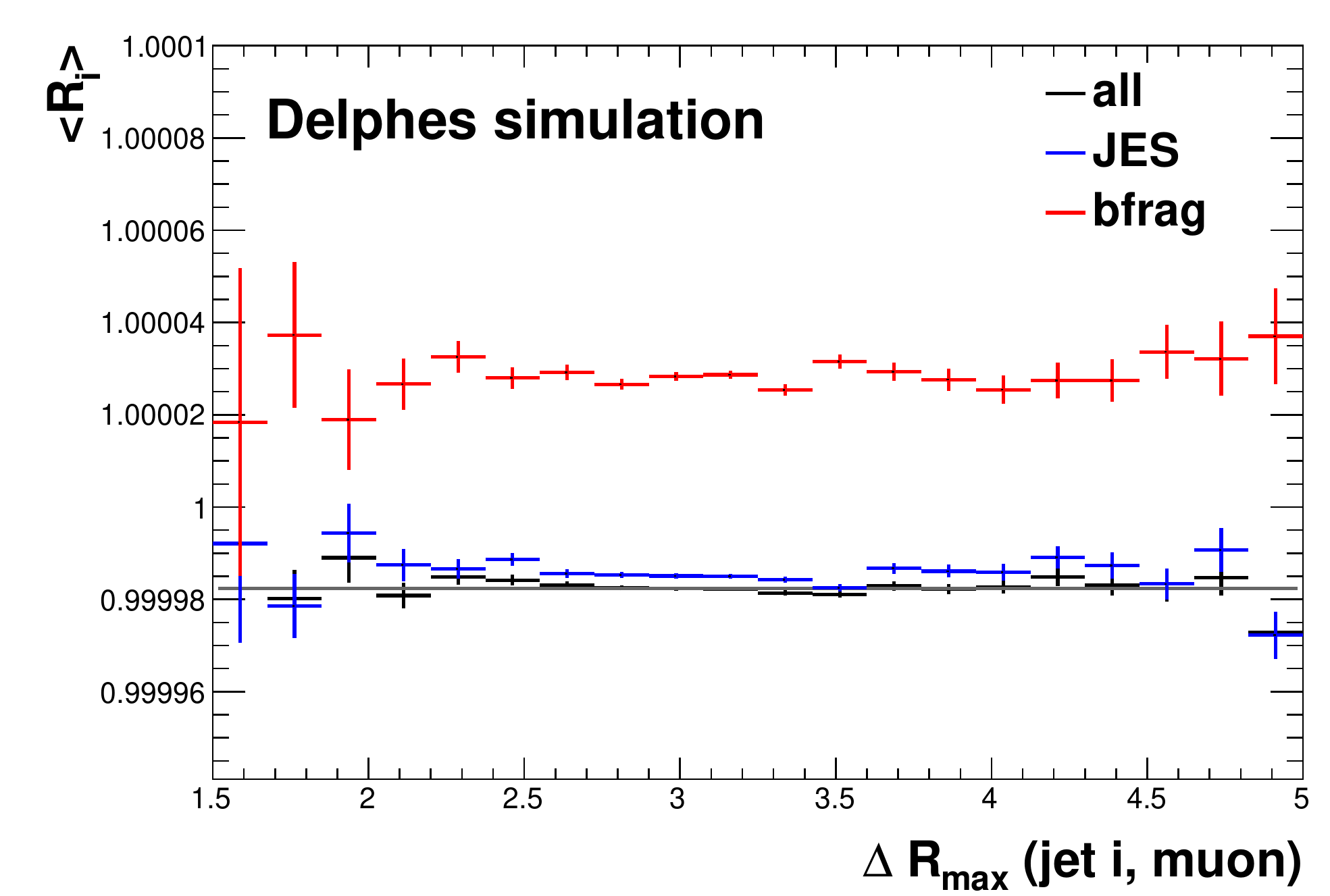}
\caption{\label{fig:vars_vs_Ri} $<R_i>$ as a function of $H_T$ (left) and as a function of the maximum $\Delta R$ between any selected jet and the muon, $\Delta R_{\mathrm{max}}(jet~i, muon)$ (right). The grey horizontal line corresponds to the overall average value of $<R_i>$. The value of $<R_i>$ for all systematic uncertainties combined (black) is driven by the two dominant systematic uncertainties: the jet energy scale and the b quark fragmentation for which the $<R_i>$ values are shown separately in blue and red, respectively. The error bars reflect the bin-by-bin uncertainty on the mean of $R_i$.}
\end{figure}

%How to use Ri to ReSyst
In equation~\ref{eq:quantifier}, the denominator has the same value for all events. When the numerator is smaller than the denominator, it means that the total systematic uncertainty becomes smaller by removing event $i$. Hence, removing events with relatively low values of $R_i$ would be a good idea to reduce the total systematic uncertainty. However, $R_i$ is an event variable that is not observable. Therefore it cannot be used directly to reject events to reduce the systematic uncertainty. Instead, the correlation between $R_i$ and event observables can be investigated to identify regions of the observable phase space that are more likely to correspond with low $R_i$ values. %From these correlations, one can try to understand why events in a certain region of the phase space have a large effect on the total systematic uncertainty or on the uncertainties induced by individual systematic sources. 
Figure~\ref{fig:vars_vs_Ri} shows the dependence of $R_i$ on the $H_T$ event observable, defined as the sum of the {\pt} of the jets in the event, and on the maximum $\Delta R$ between the muon and any of the jets in the event with $p_{\mathrm{T}}>30$~GeV. The distance $\Delta R$ is defined as $\Delta R = \sqrt{(\eta_{jet}-\eta_{muon})^2+(\phi_{jet}-\phi_{muon})^2}$, where $\eta_{jet}$ ($\eta_{muon}$) and $\phi_{jet}$ ($\phi_{muon}$) correspond respectively to the pseudorapidity and azimuthal angle of the jet (muon).
The mean of $R_i$, denoted $<R_i>$, is shown when considering all systematic sources together and separately for both the jet energy scale and b quark fragmentation uncertainties alone. It is clear that the value of $<R_i>$ is mostly driven by the impact of the jet energy scale uncertainty, which can be explained from the definition of $R_i$ in equation~\ref{eq:quantifier} and because the jet energy scale uncertainty dominates the total systematic uncertainty.
From figure~\ref{fig:vars_vs_Ri} one can also observe that $<R_i>$ can be above or below 1. This is related to the fact that the impact of the (individual) systematic effects on the top quark mass is not fully symmetric around the nominally estimated top quark mass. However, what is most important is the variation of $<R_i>$ across the observable range and the possibility to identify a region in the observable phase space corresponding to relatively lower values of $<R_i>$. In figure~\ref{fig:vars_vs_Ri}, we see that $<R_i>$ is relatively lower at small values of the $H_T$. Hence, this observation suggests to remove those events with a lower value of $H_T$ in order to reduce the total systematic uncertainty on the top quark mass measurement. 

\section{Reducing the total systematic uncertainty (ReSyst)}
\label{sec:result}
The ReSyst technique consists of applying additional event selection criteria in the event observable phase space to become less sensitive to systematic effects. Based on the behaviour of $R_i$ as a function of event observables, as for example shown in figure~\ref{fig:vars_vs_Ri}, additional selection criteria can be applied. To illustrate the power of ReSyst and motivated by the observation in figure~\ref{fig:vars_vs_Ri}, an additional event selection requirement is defined: $H_T>220$~GeV. The threshold on the $H_T$ is roughly chosen as the value where $<R_i>$ crosses the average $<R_i>$ value illustrated by the grey line. The other observable in figure~\ref{fig:vars_vs_Ri} does not exhibit significant variation in $R_i$. Other observables were also studied, but their initial dependence on $<R_i>$ disappeared after applying the additional selection criterion of $H_T>220$~GeV, as is expected for observables correlated with $H_T$. 

\begin{figure}[tbp]
\centering 
\includegraphics[width=.49\textwidth]{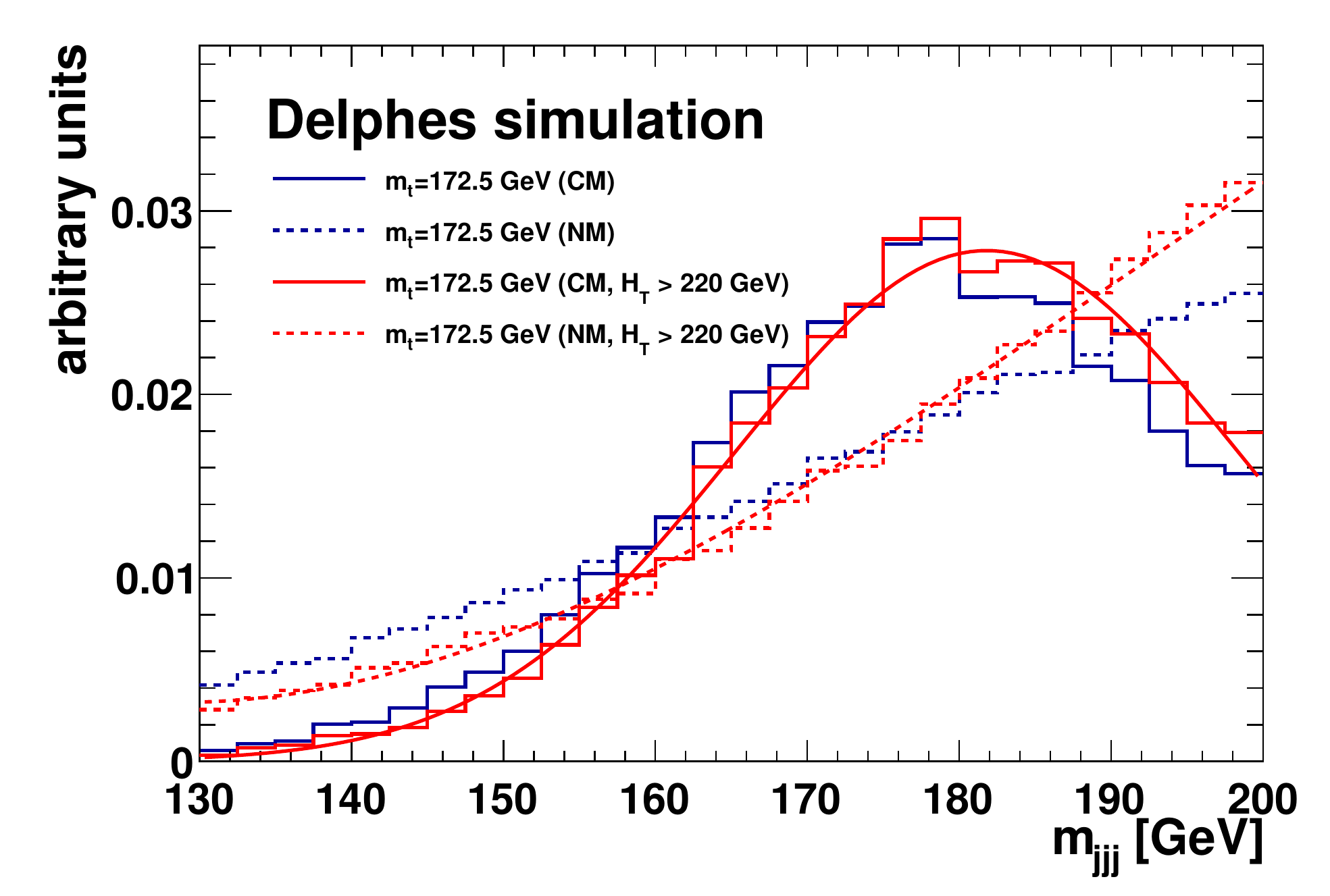}
\hfill
\includegraphics[width=.49\textwidth]{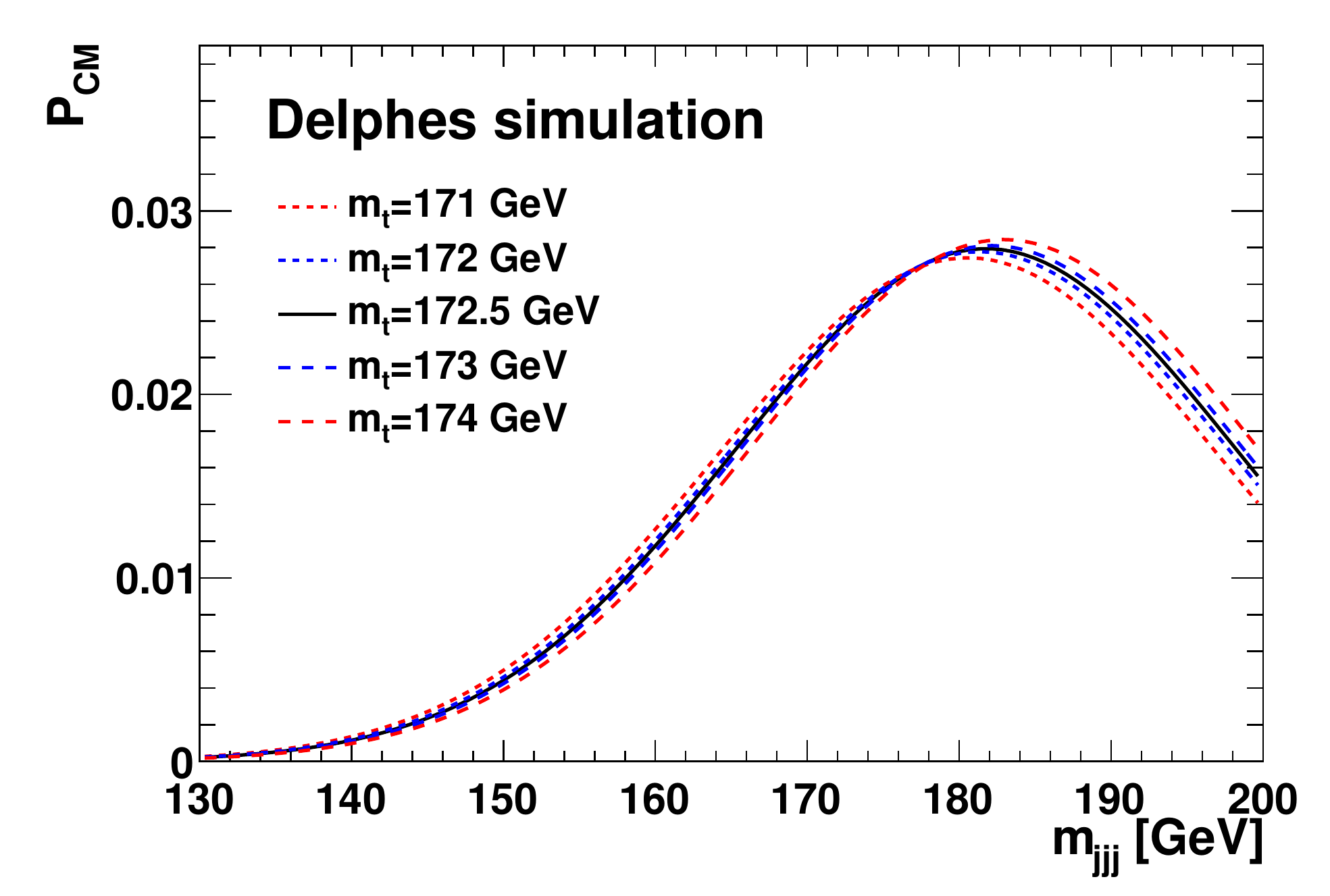}
\includegraphics[width=.49\textwidth]{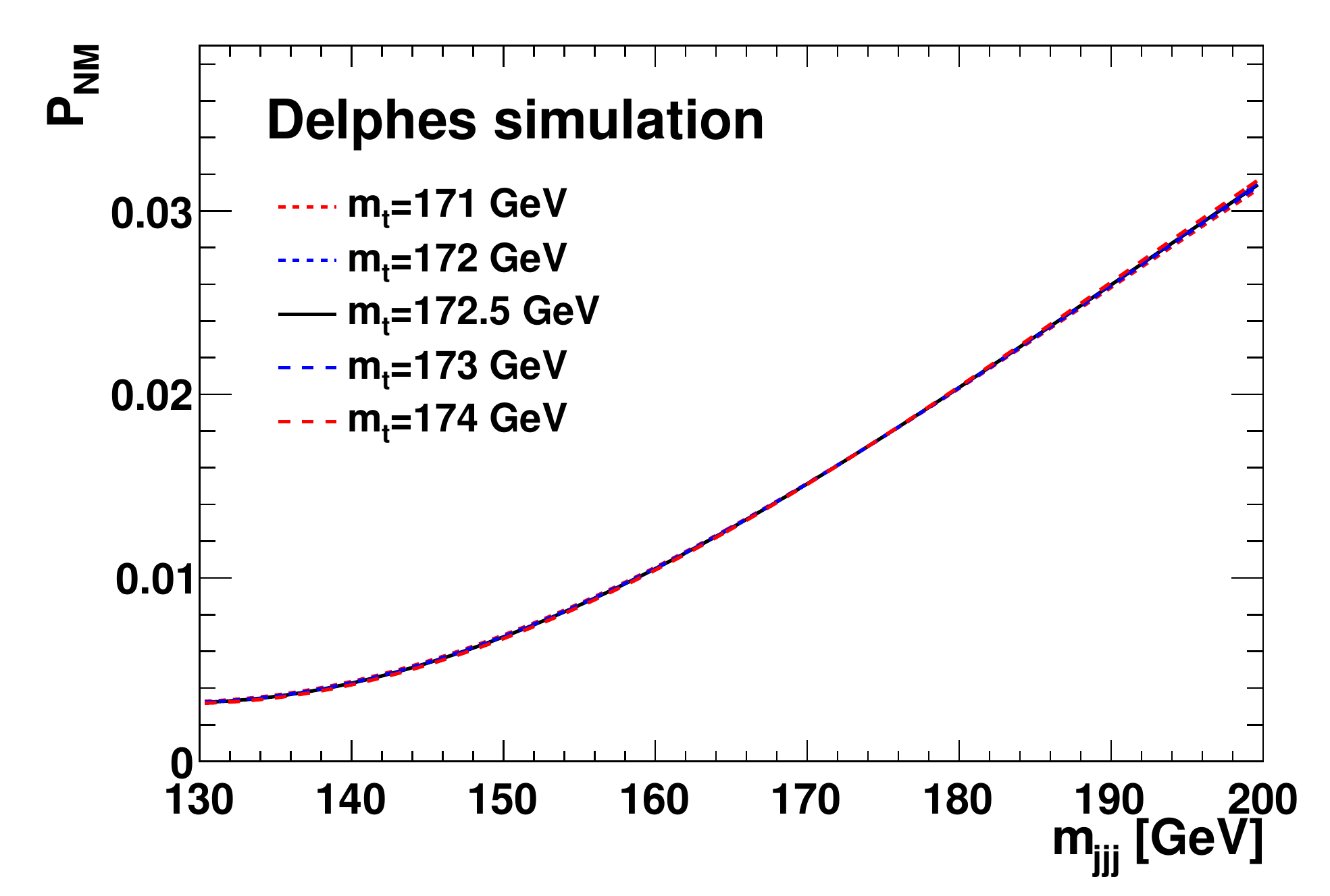}
\hfill
\includegraphics[width=.49\textwidth]{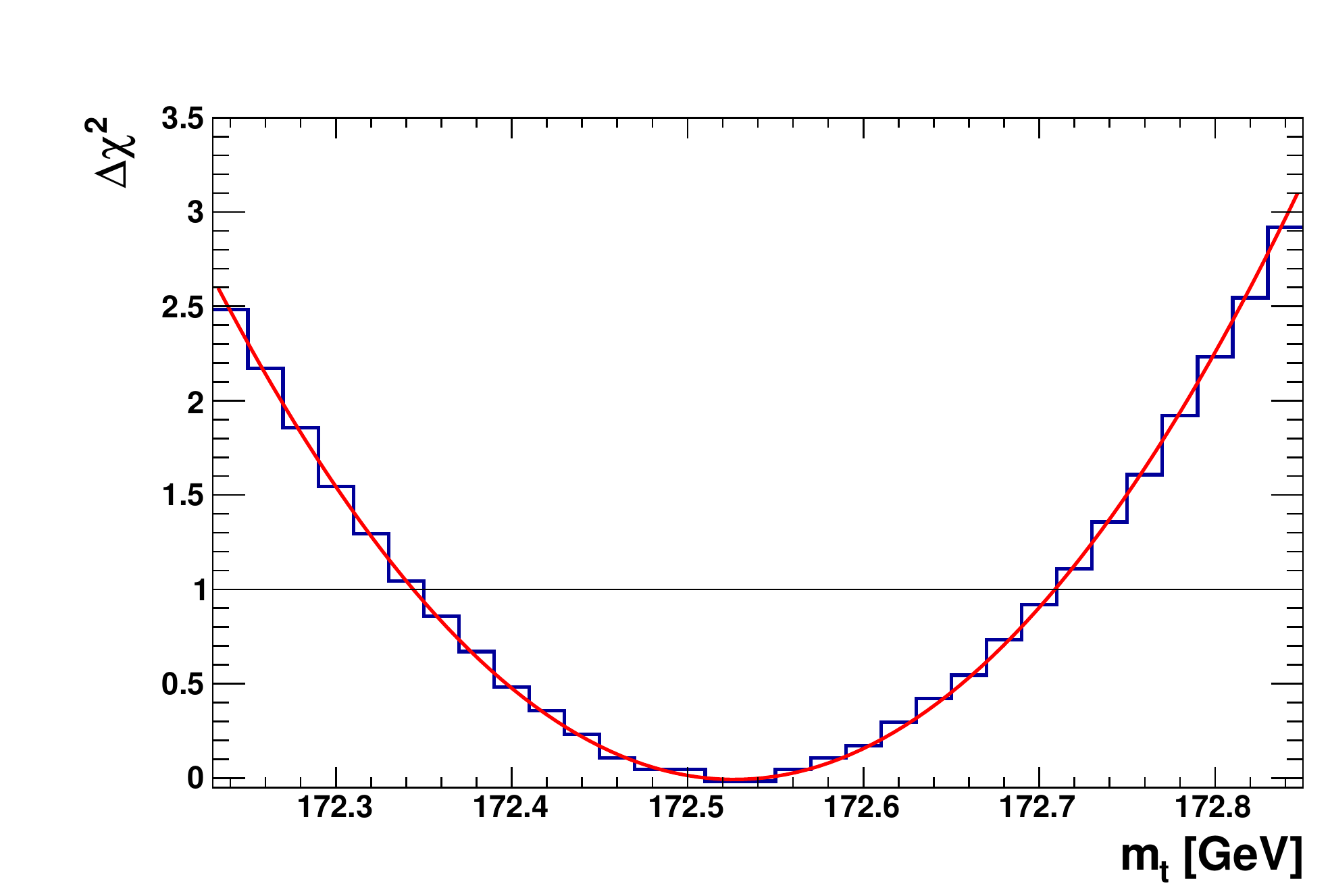}
\caption{\label{fig:afterHTcut} The same as in figure~\ref{fig:templates}, but after the additional selection requirement on $H_T$. The upper left panel also contains the distributions shown in figure~\ref{fig:templates}, i.e. before the $H_T$ requirement to illustrate how the distributions change.}
\end{figure}

After applying the additional selection requirement, an additional 31\% of the previously selected events is rejected, resulting in a total selection efficiency of 0.8\%. In contrast, the number of events for which the correct jet-quark combination is chosen raises from about 20.5\% to about 23.3\%. The probability density functions for the {\mjjj} distribution are remade with this subset of events and the obtained likelihood is maximized. Figure~\ref{fig:afterHTcut} shows the probability density functions and after the additional selection requirement. 

\begin{figure}[tbp]
\centering 
\includegraphics[width=.45\textwidth]{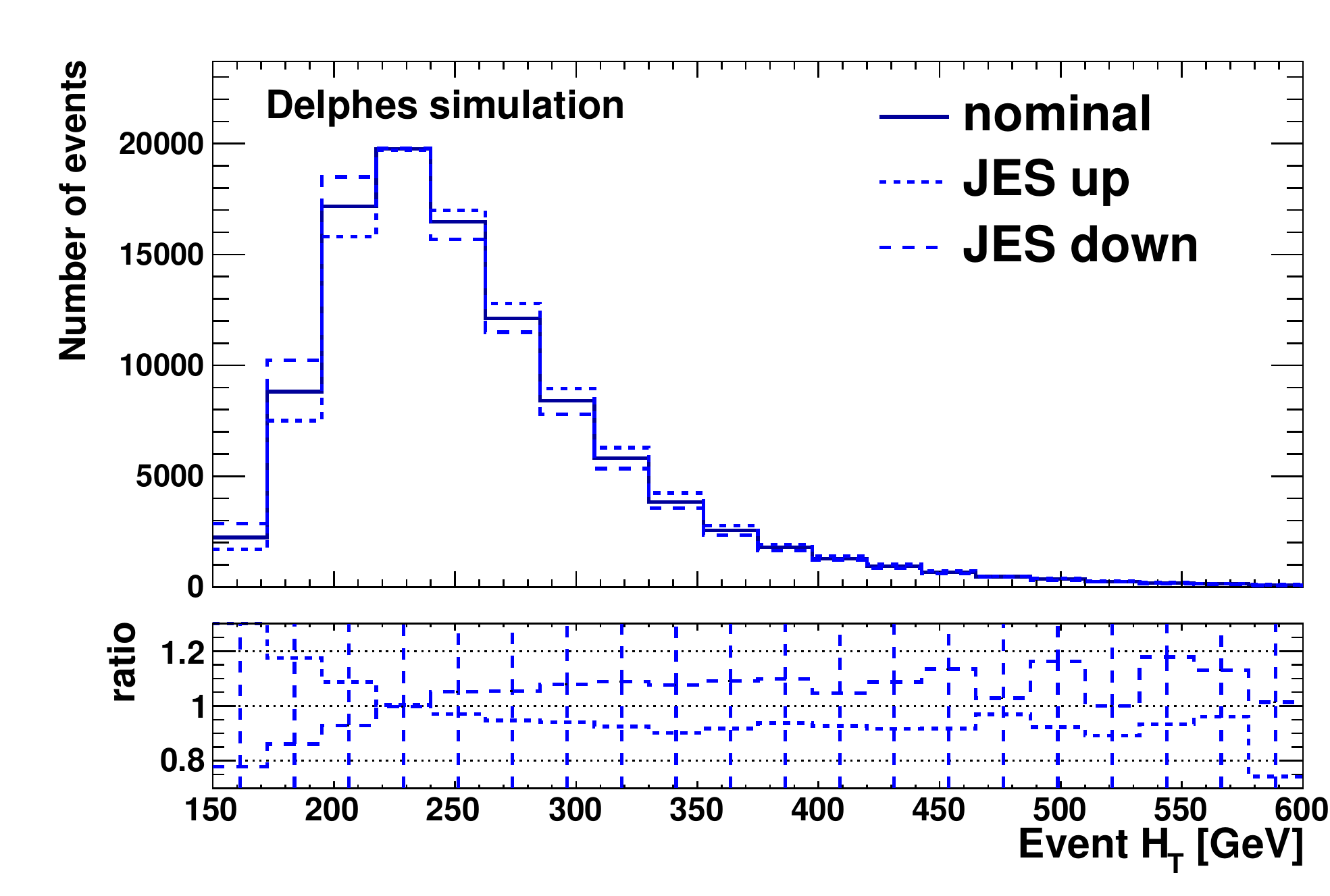}
\hfill
\includegraphics[width=.45\textwidth]{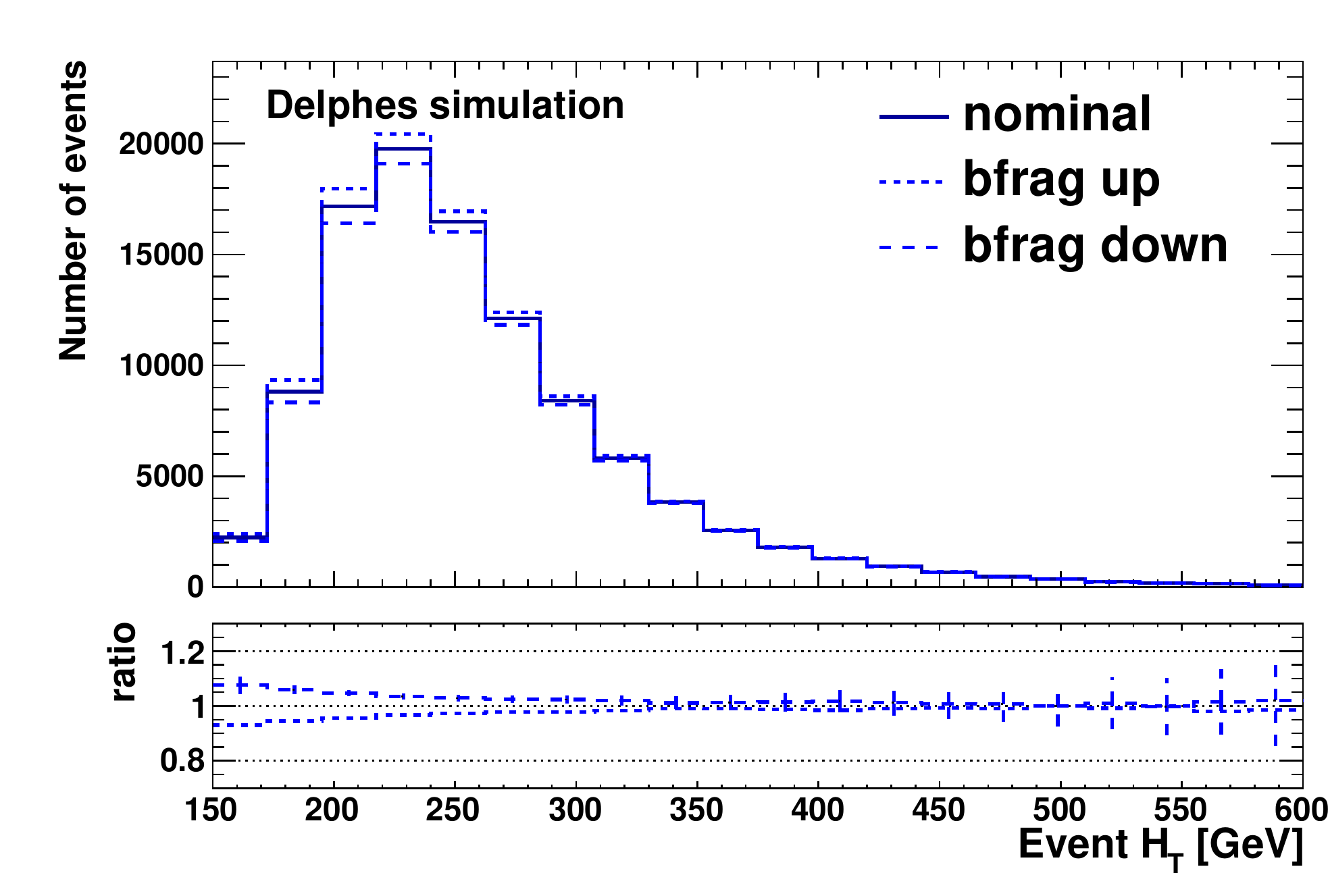}
\caption{\label{fig:jetHT} The nominal $H_T$ distribution is compared to the $H_T$ distribution obtained after varying the jet energy scale (left) and after varying the b quark fragmentation (right) according to their uncertainties. The impact of the uncertainty in the jet energy scale and in the modelling of the b quark fragmentation is indeed largest at small $H_T$ values.}
\end{figure}

The top quark mass is estimated to be $172.53 \pm 0.18~(\mathrm{stat.}) +0.67 -0.59~(\mathrm{syst.})$~GeV. 
\begin{table}[tbp]
\centering
\begin{tabular}{|l|c|c|}
\hline
Systematic source & $+1\sigma$ effect [GeV] & $-1\sigma$ effect [GeV]\\
\hline
b tagging efficiency and mistagging probability & 0.01 & -0.01 \\
Jet energy scale & 0.62 & -0.54 \\ 
Factorization and renormalization scales & 0.04 & -0.04 \\
Matrix element and parton shower matching & <0.01 & >-0.01 \\
Top quark {\pt} & 0.10 & n.a.  \\
b quark fragmentation & 0.23 & -0.23 \\
\hline
Total systematic uncertainty & 0.67 & -0.59 \\
\hline
\end{tabular}
\caption{\label{tab:systresyst}The sources of systematic uncertainties considered in the analysis after applying the $H_T>220$~GeV selection requirement proposed by the ReSyst technique and their impact on the top quark mass measurement. The uncertainties from the jet energy scale and the b quark fragmentation still dominate the total uncertainty, but their effect is reduced by about 30\% and 50\% with respect to table~\ref{tab:syst}, respectively.}
\end{table}
As expected after applying an additional selection requirement, the statistical uncertainty slightly increases. However, more important in the context of this paper is the reduction of the systematic uncertainty by about 30\% compared to the result obtained with the initial event selection discussed in section~\ref{sec:topmassest}. Table~\ref{tab:systresyst} shows the systematic uncertainties associated with the different sources. The requirement on the $H_T$ observable reduces the uncertainty due to the jet energy scale and b quark fragmentation, which are the two dominant uncertainties. Indeed, those uncertainties are expected to be reduced by the additional selection requirement since also the number of events with at least four jets may vary when the jet energy is scaled up or down, especially in case the fourth jet has a {\pt} close to 30~GeV, which is used as a threshold in the baseline selection. This can be seen in figure~\ref{fig:jetHT}, where the nominal $H_T$ distribution is compared to the distribution obtained when varying the jet energy scale and b-quark fragmentation upwards and downwards. The relative effect is indeed larger at low $H_T$ values. When applying the ReSyst technique on data, it is relevant to verify that the distributions of the chosen observables are modelled well.

For some observables the value of $<R_i>$ is found to be fairly stable across the observable range. This is the case for instance for $\Delta R_{\mathrm{max}}(jet~i, muon)$ as can be seen in the right panel of figure~\ref{fig:vars_vs_Ri}. Therefore, the systematic uncertainty is expected not to change when an additional selection requirement is placed on this observable. As an illustration, the top quark mass estimation is repeated replacing the $H_T>220$~GeV requirement with the requirement $\Delta R_{\mathrm{max}}(jet~i, muon)> 3$. About 36\% of the events are rejected by this additional requirement. To estimate the top quark mass the {\mjjj} probability density functions are remade and the top quark mass is estimated to be $173.04 \pm 0.28 ~(\mathrm{stat.}) +0.81 -0.94~(\mathrm{syst.})$~GeV. Modulo smaller changes that are due to the new {\mjjj} templates the resulting systematic uncertainty is, as expected, equivalent to the one without the additional selection criterion. This demonstrates that the ReSyst technique works conceptually.

\section{Conclusion and prospects}
\label{sec:conc}
The ReSyst technique is presented as a novel technique to guide the design of experimental analyses for precision measurements for which the statistical uncertainty is small compared to the systematic uncertainty. This technique allows for balancing statistical and systematic uncertainties by rejecting events which induce a large systematic uncertainty. A quantifier $R_i$ is introduced to assess the impact of an individual event on the systematic uncertainty. Correlating $R_i$ with event observables opens the possibility to reduce the total uncertainty. This concept is demonstrated using a simplified top quark mass estimator in the context of proton collisions at the LHC. For this estimator and for the considered systematic sources, the total systematic uncertainty is reduced by at least 30\%. This reduction is obtained without optimizing the thresholds on the observables used to reduce the total systematic uncertainty. It should be noted that a number of systematic sources were not considered, such as the number of additional pileup collisions in the same or adjacent bunch crossings, effects from the modelling of colour reconnection, or from initial and final state radiation due to the uncertainty in the strong coupling $\alpha_S$ in the parton shower. Clearly, when applying the technique on data these uncertainties should be taken into account to perform a complete top quark mass measurement. When applying additional selection requirements using the ReSyst technique, it is possible that additional systematic uncertainties need to be considered to e.g. accommodate potential differences in the selection efficiency in data and simulation.

The power of the ReSyst technique to define a quantifier for each event is at the same time also a limitation. The definition of $R_i$ assumes that systematic effects can be assessed by using the same events at the matrix-element level such that there is a one-to-one connection between the nominal event and the event processed with a different value of the parameter(s) simulating the systematic variation. For some systematic sources and generators this is not (yet) the case, e.g. for the uncertainty in the modelling of the colour reconnection in the parton shower, which is typically evaluated using an independent sample with a different generator seed.

Optimizations to the initial concept presented here are possible. For some analyses it may be better to define $R_i$ separately for the upward and downward variation of a systematic source, e.g. when the systematic uncertainty is highly asymmetric, or to define $R_i$ for a subset of systematic sources. To optimize the additional selection criteria proposed by the ReSyst method, one could perform a scan in the observable space to find the optimal thresholds on the most promising observables and/or combine observables in a multivariate analysis to find an optimal selection threshold. If a profile likelihood ratio fit is used to perform the measurement, $R_i$ could be exploited to identify control distributions and/or regions with an enhanced sensitivity on a specific systematic uncertainty source. These can be used in an overal fit to constrain the systematic uncertainties due to various sources using the data itself. Additional to quantifiying the impact of one event on the systematic uncertainty of an estimator, the statistical impact can be quantified as well, for example by considering the derivative of the likelihood at the measured value. The ReSyst method can be extended to consider both in order to minimize the total uncertainty on the estimator.

\acknowledgments
I would like to thank the Research Foundation - Flanders (FWO - Vlaanderen) for the postdoctoral fellowship. In addition, I express my deepest gratitude to the many individuals worldwide encouraging me throughout my career. This article would not exist without both forms of support.

% The bibliography will probably be heavily edited during typesetting. We'll parse it and, using the arxiv number or the journal data, will query inspire, trying to verify the data (this will probably spot eventual typos) and retrive the document DOI and eventual errata. We however suggest to always provide author, title and journal data: in short all the informations that clearly identify a document.

\end{document}